\documentstyle[aps,epsfig,psfig]{revtex}

\newcommand{\beq}{\begin{equation}}
\newcommand{\eeq}{\end{equation}}
\newcommand{\bear}{\begin{eqnarray}}
\newcommand{\ear}{\end{eqnarray}}
\newcommand{\earn}{\nonumber \end{eqnarray}}
\newcommand{\dst}{\displaystyle}
\newcommand{\nn}{\nonumber \\}
\newcommand{\Ref}[1]{(\ref{#1})}

\newcommand{\phisq}{\langle\phi^2\rangle}
\newcommand{\Tmn}{\langle T_{\mu\nu}\rangle}
\newcommand{\unren}{{\rm unren}}
\newcommand{\ren}{{\rm ren}}
\newcommand{\rc}{r_{\rm c}}
\newcommand{\mds}{m_{\rm DS}}
\newcommand{\gsim}{\mathop{\lefteqn{\raise.9pt\hbox{$>$}}\raise-3.7pt\hbox{$\sim$}}}
\newcommand{\lsim}{\mathop{\lefteqn{\raise.9pt\hbox{$<$}}\raise-3.7pt\hbox{$\sim$}}}
\arraycolsep=\mathsurround

\begin{document}

\title{Vacuum polarization of a scalar field in wormhole spacetimes}

\author{Arkadii A. Popov\thanks{e-mail: popov@kspu.kcn.ru},
Sergey V. Sushkov\thanks{e-mail: sushkov@kspu.kcn.ru}}

\address{Department of Geometry, Kazan State Pedagogical University,\\
Mezhlauk 1 st., Kazan 420021, Russia}

\maketitle

\begin{abstract}
An analitical approximation of $\langle\phi^2\rangle$ for a scalar field in
a static spherically symmetric wormhole spacetime is obtained.
The scalar field is assumed to be both massive and massless, with an arbitrary
coupling $\xi$ to the scalar curvature, and in a zero temperature vacuum state.

\vskip12pt\noindent
{\normalsize PACS number(s): 04.62.+v, 04.70.Dy}
\end{abstract}


\section{Introduction}

Wormholes are topological handles in spacetime linking widely separated
regions of a single universe, or ``bridges'' joining two different
spacetimes \cite{def}. Interest in these configurations dates back at least
as far as 1916 \cite{Flamm} with punctuated revivals of activity following
both the classic work of Einstein and Rosen in 1935 \cite{Eins} and the later
series of works initiated by Wheeler in 1955 \cite{Wheeler}. More recently, a
fresh interest in the topic has been rekindled by the works of Morris and
Thorne \cite{MT} and of Morris, Thorne and Yurtsever \cite{MTY}. These
authors constructed and investigated a class of objects they referred to as
``traversable wormholes.'' Their work led to a flurry of activity in wormhole
physics \cite{activ}.

One of the most central feature of wormhole physics is the fact that
traversable wormholes are accompanied by unavoidable violations of the null
energy condition (NEC), i.e., the matter threading the wormhole's throat has
to be possessed of ``exotic'' properties. The known classical matter does
satisfy the usual energy conditions, hence wormholes, if they exist,
should arise as solutions of general relativity and ``unusal" and/or quantum
matter.\footnote{The NEC can also be violated and wormholes can exist in
theories based on modifications of general relativity, such as Brans-Dicke
theory \cite{Agnese,Nandi,Anch}, higher derivative gravity \cite{Hoch} or
Gauss-Bonnet theory \cite{Bhawal}.} Recently, Barcel\'o and Visser
\cite{BW1,BW2} have demonstrated the possibility of existence of
traversable wormholes supported by classical non-minimally coupled scalar
fields.  In their investigation, scalar fields play a role of the
``unusual'' matter violating the NEC, and therefore, instead of the problem
of the wormhole's existence, the problem of the existence of appropriate
scalars comes to the forefront.

An alternative approach is to regard wormholes as semiclassical in nature,
and seek them as solutions of the Einstein equations, using the expectation
value of the stress-energy tensor as the source of gravity (self-consistent
wormholes). Such an idea first appeared and was realized in Ref.
\cite{Sushkov1}. Note that the main difficulty of this approach is that the
expectation values have the strong functional dependence on the metric tensor
$g_{\mu\nu}$ and are generally impossible to be calculated analitically.
For this reason, one is forced to use approximate expressions for the
expectation value of the stress-energy tensor.
In Ref. \cite{Sushkov1}, the Frolov-Zel'nikov approximation \cite{FZ} for
$\langle T_{\mu\nu} \rangle$ for conformally invariant massless fields in
static spacetimes, based on pure geometrical arguments and common properties
of the stress-energy tensor, has been used. Later, in Ref. \cite{HPS}
self-consistent spherically symmetric wormhole solutions have been found
numerically in the framework of the analytical approximation for $\Tmn$,
which was derived by Anderson, Hiscock and Samuel \cite{AHS} on the basis of
the WKB approximation for modes of a scalar field. Recently Taylor, Hiscock,
and Anderson \cite{THS}, using the Dewitt-Schwinger approximation, have
calculated analytic expressions for the stress-energy of a quantized massive
scalar field in a number of static spherically symmetric wormhole spacetimes.
As well some arguments in favour of the possibility of existence of
self-consistent wormhole solutions to semiclassical Einstein equations have
been given in Ref. \cite{Kha}.

Recently one of the authors \cite{Sushkov2} has performed approximate
calculations of $\langle\phi^2\rangle$ for a massive scalar field in static
spherically symmetric spacetimes, which improve the result by Anderson,
Hiscock, and Samuel \cite{AHS}. The improvement has been achieved due to the
use of exact expressions for mode sums and integrals. Moreover, it has been
demonstrated in Ref. \cite{Sushkov2} that the result of approximate
calculations depends on the appropriate choice of the zeroth-order
solution of the mode equation.
In this paper we apply the experience obtained in the previous work to derive
an approximate expression for $\phisq$ for a scalar field in a wormhole
spacetime.

\section{Wormhole spacetimes}

We shall regard a wormhole as a time independent, nonrotating, and
spherically symmetric bridge connecting two asymptotically flat regions.
The metric of wormhole spacetime (continued analitically into Euclidean
space) can be taken in the form, suggested by
Morris and Thorne \cite{MT}:\footnote{The units $\hbar=c=G=1$ are used
throughout the paper.}
        \beq\label{whmetric}
        ds^2=f(\rho)d\tau^2+d\rho^2
        +r^2(\rho)(d\theta^2+\sin^2\theta\, d\varphi^2).
        \eeq
Here $\tau=it$ is the Euclidean time, $\rho$ is the proper radial
distance, $\rho\in(-\infty,+\infty)$.
We assume that the redshift function $f(\rho)$ is everywhere finite (no event
horizons); the shape function $r(\rho)$ has the global minimum at $\rho=0$,
so that $r_0=\min\{r(\rho)\}=r(0)$ is the radius of the wormhole throat. The
topology of the wormhole spacetime is $R^2 \times S^2$. In order for the
spacetime geometry to tend to an appropriate asymptotically flat limit at
$\rho\to\pm\infty$ we impose
        \bear
        \lim_{\rho\to\pm\infty}\{r(\rho)/|\rho|\}=1, \quad {\rm and} \quad
        \lim_{\rho\to\pm\infty}f(\rho)=f_{\pm}.
        \ear
For simplicity we also assume symmetry under interchange of the two
asymptotically flat regions, $\rho\leftrightarrow -\rho$, that is,
$r(\rho)=r(-\rho)$ and $f(\rho)=f(-\rho)$.

Note that introducing a new radial variable $r$ instead of $\rho$ by
substitution
        \beq
        r=r(\rho)
        \eeq
allows to rewrite the metric \Ref{whmetric} in the following form
        \beq\label{metric}
        ds^2=f(r)d\tau^2+h(r)dr^2+r^2 (d\theta^2+\sin^2\theta\, d\varphi^2),
        \eeq
with
$$
h(r)=\frac1{r'^2},  
$$
where the prime means the derivative with respect to $\rho$.


\section{Approximate expression for $\phisq$}


Consider a quantized scalar field $\phi$; the scalar field is
assumed to be both massive with the mass $m$ and massless, and with an
arbitrary coupling $\xi$ to the scalar curvature.
We also assume that the field is in a vacuum
state defined with respect to the Killing vector which always
exists in a static spacetime. An unrenormalized expression for $\phisq$
can be computed from the Euclidean Green's function $G_E(x,\tilde x)$. Such
the expression given in Ref. \cite{AHS} for the case of static
spherically symmetric spacetimes reads
        \bear \label{phisq}
        \langle\phi^2(x)\rangle_{\rm unren}&=&
        G_E(x,\tau;x,\tilde\tau)
        \nonumber \\
        &=&\frac1{4\pi^2}\int_0^\infty d\omega
        \cos[\omega(\tau-\tilde\tau)]
        \sum_{l=0}^\infty \left[(2l+1)p_{\omega l}q_{\omega l}
        -\frac1{rf^{1/2}} \right],
        \ear
where the modes $p_{\omega l}$ and $q_{\omega l}$ obey the homogeneous radial
mode equation
        \bear\label{modeeqn}
        \frac{d^2S}{d\rho^2}+\left[\frac{1}{2f}\frac{df}{d\rho}
        +\frac{2}{r}\frac{dr}{d\rho}\right]\frac{dS}{d\rho}
        -\left[\frac{\omega^2}{f}+\frac{l(l+1)}{r^2}+m^2+\xi R\right]S=0.
        \ear
The modes $p_{\omega l}$ and $q_{\omega l}$ also satisfy the Wronskian
condition
        \beq\label{wronskian}
        C_{\omega l}\left[p_{\omega l}\frac{dq_{\omega l}}{d\rho}-
        q_{\omega l}\frac{dp_{\omega l}}{d\rho}\right]=-\frac1{r^2f^{1/2}},
        \eeq
where $C_{\omega l}$ is a normalization constant.

Let us stress that points in Eq. (\ref{phisqunren}) are splitted. Namely,
points are separated in time so that $\epsilon\equiv(\tau-\tilde\tau)$,
$\tilde \rho=\rho$, $\tilde \theta=\theta$, $\tilde \varphi=\varphi$. As
was first pointed out by Candelas and Howard \cite{CH} for the case of
Schwarzschild spacetime, the Euclidean Green's function have superficial
divergences with this separation of points. As discussed in Ref.
\cite{AHS}, these cannot be real divergences because the Green's
function must be finite when the points are separated; to remove the
divergences one has to subtract some additional counterterms. The terms
$(rf^{1/2})^{-1}$ in brackets in Eq. (\ref{phisqunren}) are such
counterterms (for details, see discussion in Ref. \cite{AHS}).

Now let us proceed to the WKB representation by making the change of
variables
        \beq\label{modes}
        \begin{array}{l}
        \displaystyle
        p_{\omega l}=\frac1{(2 r^2 W)^{1/2}}
        \exp\left[\int^\rho W f^{-1/2} d\rho\right], \\
        \\
        \displaystyle
        q_{\omega l}=\frac1{(2 r^2 W)^{1/2}}
        \exp\left\{-\left[\int^\rho W f^{-1/2} d\rho\right]\right\},
        \end{array}
        \eeq
where $W$ is a new function of $\rho$.
Substitution of Eq. (\ref{modes}) into Eq. (\ref{wronskian}) shows that
the Wronskian condition is obeyed if $C_{\omega l}=1$, and substitution of
Eq. (\ref{modes}) into Eq. (\ref{modeeqn}) gives the following equation
for $W$:
     \beq \label{W2}
     W^2=\omega^2+\frac{f}{r^2}l(l+1)
        +m^2 f+\xi fR
        +\frac{r'f'}{2r}+\frac{r''f}{2r}
        +\frac{f'}{4}\frac{W'}{W}+\frac{f}{2}\frac{W''}{W}
        -\frac{3f}{4}\frac{W'^2}{W^2}
     \eeq
with
        \beq
        R=-\frac{f''}{f}+\frac{f'^2}{2f^2}-\frac{2r'f'}{rf}
          -\frac{4r''}{r}-\frac{2r'^2}{r^2}+\frac{2}{r^2}.
        \eeq
Assuming that the functions $f$ and $r$ are varying sufficiently slowly, one
can solve Eq. \Ref{W2} iteratively with the small parameter
     \beq
     \lambda_{WKB}=L_{\star} /L \ll 1,
     \eeq
where $L$ is a characteristic scale of varying of the metric
functions\footnote{More exactly, one may define $L$ as follows:
$L^{-1}= \max \left \{ \left| \left[ \ln(fr) \right]'  \right|, \
       \left| \left[ \ln(fr) \right]''  \right|^{1/2}, \
       \left| \left[ \ln(fr) \right]'''  \right|^{1/3}, \
       \dots  \right \}$.}
and
      \beq
      L_{\star} =\left[ m^2+ \frac{2\xi}{r^2} \right]^{-1/2}
      \eeq
Neglecting terms with derivatives in Eq. \Ref{W2} we choose the zeroth-order
solution as follows: $$W=\Omega,$$ where
        \bear
        \Omega&=&\left[\omega^2+\frac{f}{r^2} l(l+1)
        +f m^2+\frac{2\xi}{r^2}\right]^{1/2}
\nn     &=&\left[\omega^2+\frac{f}{r^2}\left(l+\frac12\right)^2
        +\frac{f}{r^2}\mu^2\right]^{1/2}.
        \ear
Here we denote
      \beq
      \mu^2=m^2r^2+2\delta
      \eeq
with $\delta=\xi-1/8$. Stress that below we assume $\mu^2>0$.
The fourth-order solution is
        \beq\label{wkbsol}
        W=\Omega+W^{(2)}+W^{(4)}
        \eeq
with
        \beq
        W^{(2)}=\frac12\frac{V}{\Omega}+\frac{f}{4}\frac{\Omega''}{\Omega^2}
        +\frac{f'}{8}\frac{\Omega'}{\Omega^2}
        -\frac{3f}{8}\frac{{\Omega'}^2}{\Omega^3}
        \eeq
and
        \beq
        W^{(4)}=-\frac{{W^{(2)}}^2}{2\Omega}
        +\frac{f}{4}\frac{{W^{(2)}}''}{\Omega^2}
        +\frac{f'}{8}\frac{{W^{(2)}}'}{\Omega^2}
        -\frac{f}{4}\frac{\Omega''W^{(2)}}{\Omega^3}
        -\frac{f'}{8}\frac{\Omega'W^{(2)}}{\Omega^3}
        +\frac{3f}{4}\frac{{\Omega'}^2 W^{(2)}}{\Omega^4},
        \eeq
where
        \beq
        V=\frac{r'f'}{2r}+\frac{r''f}{2r}+\xi f\left(R-\frac2{r^2}\right).
        \eeq

Substituting the solution \Ref{wkbsol} into \Ref{modes} and \Ref{phisqunren},
and neglecting terms of the sixth order and higher we can obtain the
following expression for the fourth-order WKB approximation for $\phisq$:
     \bear \label{phisqunren}
     \phisq_\unren&=&\frac{1}{4\pi^2 r^2}\int_0^\infty
     d\omega\cos[\omega(\tau-\tilde\tau)]\sum_{l=0}^\infty
     \left({l+\frac12}\right) \left\{ \left[\frac{1}{\Omega}
     -\frac{r}{f^{1/2}(l+\frac12)} \right]
     -\frac{V}{2\Omega ^3} \right. \nn
     &&+\frac{f^2}{\Omega ^5} \left[ -\frac{U''}{8f}
     -\frac{f'U'}{16f^2}+\frac{3V^2}{8f^2}-\frac{f'V'}{16f^2}
     -\frac{V''}{8f}  \right]
     +\frac{f^2}{\Omega^7} \left[
      \frac{5 U'^2}{32f}-\frac{f''U''}{16f}
     -\frac{3f'^2 U''}{128f^2}
     \right.\nn &&\left.
     +\frac{5VU''}{16f}
     -\frac{f'''U'}{64f}
     -\frac{f'f''U'}{128f^2} +\frac{5V'U'}{16f}
     +\frac{5f'VU'}{32f}
     - \frac{3f'U'''}{32f}
     -\frac{U''''}{32}  \right] \nn
     &&+\frac{f^2}{\Omega^9} \left[
     \frac{7f''U'^2}{64f}
     +\frac{21 f'^2U'^2}{512f^2}
     -\frac{35VU'^2}{64f}
     +\frac{63f'U'U''}{128f}
     +\frac{7U'U'''}{32}
     +\frac{21U''^2}{128}\right] \nn
     &&+\frac{f^2}{\Omega^{11}} \left[
     -\frac{231f'U'^3}{512f}
     -\frac{231U'^2U''}{256}\right]
     \left.
     +\frac{f^2}{\Omega^{13}}
     \frac{1155U'^4}{2048}
     \right\},
     \ear
where
        \beq
        U=\frac{f}{r^2}\left(l+\frac12\right)^2
          +f\left(m^2+\frac{2\delta}{r^2}\right).
        \eeq

A renormalized expression for $\phisq$ can be obtained as follows:
        \beq\label{subtr}
        \phisq_{\rm ren}=\lim_{\epsilon\to 0}
        \left(\phisq_{\rm unren}-\phisq_{\rm DS}\right),
        \eeq
where the renormalization counterterms for a massive scalar field are given
by \cite{Chris}
        \bear
        \phisq_{\rm DS} &=& G_{\rm DS}(x,x') \nonumber \\
        &=&\frac1{8\pi^2\sigma}+\frac1{8\pi^2}
        \left[m^2+\left(\xi-\frac16\right)R\right]
        \left[C+\frac12\ln\left( \frac{\mds^2|\sigma|}{2}
        \right)\right] -\frac{m^2}{16\pi^2}
        +\frac1{96\pi^2}R_{\alpha\beta}
        \frac{\sigma^\alpha\sigma^\beta}{\sigma}.
        \ear
Here $\sigma$ is equal to one half the square of the distance between the
points $x$ and $x'$ along the shortest geodesic connecting them. $C$ is
Euler's constant, $R_{\alpha\beta}$ is the Ricci tensor and
$\sigma^\alpha\equiv\sigma^{;\alpha}$. The constant $\mds$ is equal to the
mass $m$ of the field for a massive scalar field. For a massless scalar field
it is an arbitrary parameter. A particular choice of the value of $\mds^2$
corresponds to a finite renormalization of the coefficients of terms in the
gravitational Lagrangian.

To perform the procedure of renormalization in practice we make use of the
formulas obtained in Ref. \cite{Sushkov2}. These formulas give the
expressions for mode sums and integrals as exact expansions in powers of
$\epsilon$:
        \bear\label{S}
        S_0(\epsilon,\mu)&=&\int_0^\infty du
        \cos\left(u\frac{\epsilon f^{1/2}}{r}\right)
        \sum_{l=0}^\infty
        \left[\frac{l+1/2}{\sqrt{u^2+\mu^2+(l+1/2)^2}}-1\right] \nn
        &=&\frac{r^2}{f\epsilon^2}
        +\frac12\left(\mu^2-\frac{1}{12}\right)
        \left(\ln \frac{\epsilon\mu f^{1/2}}{2r}+C\right)
        -\frac{\mu^2}{4}-\mu^2 S_0(2\pi\mu)
        +{\rm O}(\epsilon^2\ln\epsilon),
        \\[12pt]
        S_n(\epsilon,\mu)&=&\int_0^\infty du
        \cos\left(u\frac{\epsilon f^{1/2}}{r}\right)
        \sum_{l=0}^\infty
        \frac{(l+1/2)^{2n-1}}{[u^2+\mu^2+(l+1/2)^2]^{n+1/2}},
        \nn
        &=&-\frac{2^{n-1}(n-1)!}{(2n-1)!!}
        \left(\ln\frac{\epsilon\mu f^{1/2}}{2r}+C\right)
        +\frac{S_n(2\pi\mu)}{(2n-1)!!}
        +{\rm O}(\epsilon^2\ln\epsilon),
        \nn
        \lefteqn{n=1,2,3,\dots,}\phantom{S_n(\epsilon,\mu)},
        \nonumber
        \ear
where $(2n-1)!!\equiv 1\cdot3\cdot5\cdots(2n-1)$ and
        \beq
        \begin{array}{l}
        \dst
        S_0(2\pi\mu)=\int_0^\infty\frac{x\ln|1-x^2|}
        {e^{2\pi\mu x}+1}dx, \\[12pt]
        \dst
        S_n(2\pi\mu)=\int_0^\infty dx \ln|1-x^2| \frac{d}{dx}
        \left(\frac1x\frac{d}{dx}\right)^{n-1}
        \frac{x^{2(n-1)}}{e^{2\pi\mu x}+1}.
        \end{array}
        \eeq
Together with the integrosums \Ref{S}, which diverge in the limit
$\epsilon\to 0$, the expression \Ref{phisqunren} contains the
integrosums of the following form:
        \bear\label{Nepsilon}
        N_n^m(\epsilon,\mu)&=&\int_0^\infty du
        \cos\left(u\frac{\epsilon f^{1/2}}{r}\right)
        \sum_{l=0}^\infty
        \frac{(l+1/2)^m}{[u^2+\mu^2+(l+1/2)^2]^{n+1/2}},
        \\
        \lefteqn{n=2,3,4,\dots,}\phantom{N_n^m(\epsilon,\mu)}
        \nonumber \\
        \lefteqn{m=1,3,\dots,2n-3}\phantom{N_n^m(\epsilon,\mu)}
        \nonumber
        \ear
The functions $N_n^m(\epsilon,\mu)$ are regular in the limit
$\epsilon\to 0$, and so one may directly set $\epsilon=0$ in Eq.
\Ref{Nepsilon}. Then, changing the order of summation and integration in
Eq. \Ref{Nepsilon} and integrating over $u$ gives
        \beq\label{N}
        N_n^m(0,\mu)=\frac{(2n-2)!!}{(2n-1)!!} \, N_n^m(\mu),
        \eeq
with
      $$
       N_n^m(\mu)=\sum_{l=0}^\infty
        \frac{(l+\frac12)^m}{[\mu^2+(l+\frac12)^2]^n}.
      $$

Substituting Eqs. (\ref{S},\ref{N}) into Eq. (\ref{phisqunren}) we may
find the asymptotical expansion for $\phisq_{\rm unren}$ in the limit
$\epsilon\to 0$. Finally, carring out the procedure of renormalization
(\ref{subtr}), i.e. subtracting $\phisq_{\rm DS}$, we obtain the
renormalized expression for $\phisq$ in the framework of the fourth-order
WKB approximation:
     \bear \label{phi2}
     4\pi^2 && \phisq_{\ren}  =
     \frac14\left[m^2+\left(\xi-\frac 16\right)R\right]\,
     \ln\left(\frac{\mu^2}{\mds^2 r^2}\right)
     -\frac{\delta}{2r^2}
     -\frac{\mu^2}{r^2}\,S_0(2\pi\mu)
     +\frac{f''}{24f}-\frac{f'^2}{24f^2}+\frac{f'r'}{12fr}
   \nn
     && -\frac{V}{2f} S_1(2\pi\mu)
     -S_2(2\pi\mu)\frac{r^2}{24f^2}\Bigg[f\left(\frac{f}{r^2}\right)''
     +\frac12 f'\left(\frac{f}{r^2}\right)'
     \Bigg]
     + S_3(2\pi\mu)\frac{r^4}{96f^2}\left(\frac{f}{r^2}\right)'^2
   \nn
     &&- N_2^1(\mu)\frac{r^2}{12f}\left\{
     \left({\frac{f}{r^2}\mu^2}\right)''
     +\frac12 \frac{f'}{f}\left({\frac{f}{r^2}\mu^2}\right)'
     -\frac{3 V^2}{f}+ V''+\frac12\frac{f'}{f}V' \right\}
   \nn
     &&+N_3^1(\mu)\frac{r^4}{12f^2}\left\{
     \left(\frac{f\mu^2}{r^2}\right)'^2
     -\left(\frac{\mu^2f}{r^2}\right)'
     \left(\frac{1}{10} f'''+
     \frac{1}{20}\frac{f'f''}{f}-2V'-
     \frac{f'V}{f} \right) \right.
   \nn
     && \left. -\left(\frac{\mu^2f}{r^2}\right)'' \left(
     \frac{2}{5}f''+\frac{3}{20}\frac{f'^2}{f}-2V \right)
     -\frac35\frac{f'}{f}\left(\frac{\mu^2f}{r^2}\right)'''
     -\frac15\left(\frac{\mu^2f}{r^2}\right)''''
     \right\}
  \nn
     &&+N_3^3(\mu)\frac{r^4}{12f^2}\left\{2\left(\frac{f}{r^2}\right)'
     \left(\frac{\mu^2f}{r^2}\right)'
  \right.
     -\left(\frac{f}{r^2}\right)'\left(
      \frac{1}{10}f'''+\frac{1}{20}\frac{f'f''}{f}-2V'-\frac{f'V}{f}\right)
  \nn
     && \left. -\left(\frac{f}{r^2}\right)''\left(
     \frac{2}{5}f''+\frac{3}{20}\frac{f'^2}{f}-2V \right)
     -\frac35\frac{f'f'''}{f}-\frac15f'''' \right\}
  \nn
     &&+N_4^5(\mu)\frac{r^6}{10f^3}\left\{\left(\frac{f}{r^2}\right)'^2
     \left(\frac{1}{2}f''+\frac{3}{16}\frac{f'^2}{f}-\frac52 V\right)
     +\frac{9}{4}f'\left(\frac{f}{r^2}\right)'\left(\frac{f}{r^2}\right)''
     \right.
  \nn
     &&\left.
     +f\left(\frac{f}{r^2}\right)'\left(\frac{f}{r^2}\right)'''
     +\frac{15}{28}f\left(\frac{f}{r^2}\right)''^2\right\}
  \nn
     &&+N_4^3(\mu)\frac{r^6}{10f^3}\left\{\left(\frac{f}{r^2}\right)'
     \left(\frac{\mu^2f}{r^2}\right)'\left(f''
     +\frac{3}{8}\frac{f'^2}{f}-5V\right)
     +\frac{9}{4}f'\left(\frac{f}{r^2}\right)'
     \left(\frac{\mu^2f}{r^2}\right)''
     \right.
  \nn
     && \left. +\frac{9}{4}f'\left(\frac{f}{r^2}\right)''
     \left(\frac{\mu^2f}{r^2}\right)'
     +f\left(\frac{f}{r^2}\right)'
     \left(\frac{\mu^2f}{r^2}\right)'''+
     f\left(\frac{f}{r^2}\right)'''\left(\frac{\mu^2f}{r^2}\right)'
     +\frac{15}{14}f\left(\frac{f}{r^2}\right)''
     \left(\frac{\mu^2f}{r^2}\right)'' \right\}
  \nn
     &&+N_4^1(\mu)\frac{r^6}{10f^3}\left\{
     \left(\frac{\mu^2f}{r^2}\right)'^2
     \left( \frac12 f''+\frac{3}{16}\frac{f'^2}{f}-\frac52V \right)
     +\frac{9}{4}f'\left(\frac{\mu^2f}{r^2}\right)'
     \left(\frac{\mu^2f}{r^2}\right)''
\right.  \nn
     && \left. +f\left(\frac{\mu^2f}{r^2}\right)'
     \left(\frac{\mu^2f}{r^2}\right)'''
     +\frac{3}{4}f\left(\frac{\mu^2f}{r^2}\right)''^2
     \right\}
     -N_5^7(\mu)\frac{11r^8}{30f^4}\left\{\frac12 f'
     \left(\frac{f}{r^2}\right)'^3+f\left(\frac{f}{r^2}\right)'^2
     \left(\frac{f}{r^2}\right)'' \right\}
  \nn
     &&-N_5^5(\mu)\frac{11r^8}{30f^4}\left\{\frac32 f'
     \left(\frac{f}{r^2}\right)'^2\left(\frac{\mu^2f}{r^2}\right)'
     +2f\left(\frac{f}{r^2}\right)'
     \left(\frac{f}{r^2}\right)''\left(\frac{\mu^2f}{r^2}\right)'
     +f\left(\frac{f}{r^2}\right)'^2
     \left(\frac{\mu^2f}{r^2}\right)''\right\}
  \nn
     &&-N_5^3(\mu)\frac{11r^8}{30f^4}\left\{ \frac32 f'
     \left(\frac{f}{r^2}\right)'\left(\frac{\mu^2f}{r^2}\right)'^2
     +f\left(\frac{f}{r^2}\right)''\left(\frac{\mu^2f}{r^2}\right)'^2
     +2f\left(\frac{f}{r^2}\right)'
     \left(\frac{\mu^2f}{r^2}\right)'\left(\frac{\mu^2f}{r^2}\right)''
     \right\}
  \nn
     &&-N_5^1(\mu)\frac{11r^8}{30f^4}\left\{\frac12 f'
     \left(\frac{\mu^2f}{r^2}\right)'^3
     +f\left(\frac{\mu^2f}{r^2}\right)'^2
     \left(\frac{\mu^2f}{r^2}\right)'' \right\}
  \nn
     &&+N_6^9(\mu)\frac{5r^{10}}{24f^4}\left(\frac{f}{r^2}\right)'^4
     +N_6^7(\mu)\frac{5r^{10}}{6f^4}\left(\frac{f}{r^2}\right)'^3
      \left(\frac{\mu^2f}{r^2}\right)'
     + N_6^5(\mu)\frac{5r^{10}}{4f^4}\left(\frac{f}{r^2}\right)'^2
      \left(\frac{\mu^2f}{r^2}\right)'^2
 \nn
     &&+N_6^3(\mu)\frac{5r^{10}}{6f^4}\left(\frac{f}{r^2}\right)'
      \left(\frac{\mu^2f}{r^2}\right)'^3
     +N_6^1(\mu)\frac{5r^{10}}{24f^4}\left(\frac{\mu^2 f}{r^2}\right)'^4.
     \ear


\section{Analysis}

The analytical approximation for $\phisq$, given by Eq. \Ref{phi2}, may be
separately analysed for two different cases: (i) $r(\rho)\gsim L$, and (ii)
$r(\rho)\ll L$.

%
\subsection{The case $r\gsim  L$}
%
First, let us consider $\phisq$ in the region $r(\rho)\gsim L$. Note that in
this case $\mu^2\approx m^2r^2\gg 1$. Really, it immediately follows from the
basic WKB condition $\lambda_{WKB}=L_\star/L\ll 1$:
      $$
      1 \ll \frac{L^2}{L_\star^2}=L^2\left(m^2+\frac{2\xi}{r^2}\right)
      \lsim r^2\left(m^2+\frac{2\xi}{r^2}\right)=m^2r^2+2\xi,
      $$
assuming that $\xi$ is of order unity or less, we conclude that
$m^2r^2 \gg 1$.
The functions $S_n(2\pi\mu)$ and $N_n^m(\mu)$ have a
simple asymptotical form at large values of $mr$; the respective
formulas are given in Appendix (see Eqs. \Ref{asymS}, \Ref{asymN}).
Substituting these formulas into Eq. \Ref{phi2} and taking into account
the asymptotical expansion $\ln(1+x)\approx
x-\frac12 x^2$ gives
      \bear \label{phi2A}
      4\pi^2 && m^2 \phisq_\ren =
      \frac{5\xi-1}{120}\frac{f''''}{f}
      -\frac{5\xi-1}{48}\frac{f'''f'}{f^2}
      +\left(-\frac18\xi^2+\frac{7}{24}\xi-\frac{13}{240}\right)
            \frac{f''f'^2}{f^3}
\nn
    &&+\left(\frac18\xi^2-\frac18\xi+\frac{23}{1680}\right)\frac{f''^2}{f^2}
      +\left(\frac1{32}\xi^2-\frac{11}{96}\xi+\frac7{320}\right)
            \frac{f'^4}{f^4}
\nn
    &&-\left(\frac12\xi^2-\frac16\xi+\frac1{72}\right)\frac{f''}{fr^2}
      +\left(\frac14\xi^2-\frac1{12}\xi+\frac1{144}\right)\frac{f'^2}{f^2r^2}
\nn
    &&+\frac{5\xi-1}{30}\frac{f'''r'}{fr}
      +\left(\frac12\xi^2-\frac{13}{24}\xi+\frac{73}{560}\right)
            \frac{f''f'r'}{f^2r}
      +\left(\xi^2-\frac1{24}\xi-\frac{13}{1680}\right)\frac{f''r''}{fr}
\nn
    &&+\left(\frac12\xi^2-\frac16\xi-\frac{127}{2520}\right)
            \frac{f''r'^2}{fr^2}
      -\left(\frac14\xi^2-\frac7{24}\xi+\frac{23}{480}\right)
            \frac{f'^3r'}{f^3r}
      -\left(\frac12\xi^2+\frac1{48}\xi-\frac{11}{480}\right)
            \frac{f'^2r''}{f^2r}
\nn
    &&+\left(\frac14\xi^2-\frac18\xi-\frac{83}{2016}\right)
            \frac{f'^2r'^2}{f^2r^2}
      +\left(\frac16\xi-\frac{11}{480}\right)\frac{f'r'''}{fr}
      +\left(2\xi^2-\frac12\xi-\frac{401}{10080}\right)\frac{f'r''r'}{fr^2}
\nn
    &&+\left(\xi^2-\frac{5}{12}\xi+\frac{179}{840}\right)\frac{f'r'^3}{fr^3}
      -\left(\xi^2-\frac5{12}\xi+\frac1{24}\right)\frac{f'r'}{fr}
      +\left(\frac16\xi-\frac1{80}\right)\frac{r''''}{r}
\nn
    &&+\frac{5\xi-1}{30}\frac{r'''r'}{r^2}
      +\left(2\xi^2-\frac{11}{12}\xi+\frac{163}{840}\right)
            \frac{r''r'^2}{r^3}
      +\left(2\xi^2-\frac16\xi-\frac{97}{3360}\right)\frac{r''^2}{r^2}
\nn
    &&+\left(\frac12\xi^2-\frac{61}{420}\right)\frac{r'^4}{r^4}
      -\left(2\xi^2-\frac56\xi+\frac7{96}\right)\frac{r''}{r^3}
      -\left(\xi^2-\frac16\xi\right)\frac{r'^2}{r^4}
\nn
    &&+\left(\frac12\xi^2-\frac16\xi+\frac1{60}\right)\frac{1}{r^4}.
      \ear

The expression for $\phisq$, given by Eq. \Ref{phi2A}, has same order as the
term proportionate to $m^{-2}$ in DeWitt-Schwinger expansion for $\phisq$,
however it does not coincide with them. The reason of this is that the
topological structures of Minkowski and wormhole spacetimes are different.

As a simple example, we consider the wormhole spacetime with the metric
        \beq\label{metricMT}
        ds^2=-dt^2+d\rho^2
            +(\rho^2+r_0^2)(d\theta^2+\sin^2\theta\, d\varphi^2),
        \eeq
given by Morris and Thorne \cite{MT}. Here $f(\rho)\equiv 1$ and
$r(\rho)=(\rho^2+r_0^2)^{1/2}$. The
characteristic scale of varying of the metric functions can be defined as
follows: $L^{-2}\sim r''/r=r_0^2(\rho^2+r_0^2)^{-2}$ near the throat, and
$L^{-1}\sim r'/r=\rho(\rho^2+r_0^2)^{-1}$ far from the throat. Stress
that in both cases $r(\rho)\sim L$. Note also that the basic WKB condition
$L/L_\star\gg 1$ is satisfied provided $r_0/\rc\gg 1$, where
$\rc=m^{-1}$ is the Compton length. Carrying out appropriate calculations in
Eq. \Ref{phi2A} we find
      \bear \label{phi2MT}
      4\pi^2m^2 \phisq_\ren&=&\frac{\Psi_\xi(x)}{r_0^4}
      \ear
where
      \beq
      \Psi_\xi(x)=-\frac{1}{(x^2+1)^4}\left(\frac9{70}x^4-
      \frac{117}{1120}x^2-\frac54\xi x^2-\frac12\xi^2+\frac1{21}\right)
      \eeq
and $x=\rho/r_0$ is a dimensionless proper radial distance. The obtained
formula obviously shows that the qualitative behavior of $\phisq$ is
determined by the function $\Psi_\xi(x)$ and depends only on $\xi$.
This dependance is very interesting. In Figs. 1,2 we present
the family of curves of $\phisq$ for various values of $\xi$.
It is seen that $\phisq$ is purely negative at $\xi=0$, and
becomes alternating for greater values of $|\xi|$. To characterize this
feature more accurate we consider an averaged value of
$\phisq$, i.e.,
      \beq
      \int_{-\infty}^{\infty}\phisq\,r(\rho)d\rho=
      \frac{1}{4\pi^2m^2r_0^2}
      \int_{-\infty}^\infty\Psi_\xi(x)\sqrt{x^2+1}dx=
      \frac{1}{4\pi^2m^2r_0^2}\left(
      \frac8{15}\xi^2+\frac13\xi-\frac{937}{12600}\right).
      \eeq
It is easily to find that the averaged value is negative provided
$\xi_1<\xi<\xi_2$, where $\xi_1\approx -0.7994$ and $\xi_2\approx 0.1744$,
and positive in the other cases. At last, we note that the average value of
$\phisq$, computed in the metric \Ref{metricMT}, is negative for $\xi=0$
(minimal coupling) and $\xi=1/6$ (conformal coupling).

%
\subsection{The case $r\ll L$}
%
Now let us analyse the analytical approximation \Ref{phi2} in the region
$r(\rho)\ll L$. Remember that $L$ is the characteristic scale of varying of
the metric, and so $L^{-1}$ is proportional to derivatives of the
metric functions. Hence it follows from the relation $r(\rho)\ll L$
that one may neglect terms with derivatives in Eq. \Ref{phi2}
with respect to terms without ones. As the result, we obtain
      \bear \label{phi2B}
      4\pi^2\phisq_{\ren} &=&
      \frac14\left[m^2+\frac{2}{r^2}\left(\xi-\frac 16\right) \right]
     \ln\left(\frac{\mu^2}{\mds^2 r^2}\right)
     -\frac{\delta}{2r^2}
     -\frac{\mu^2}{r^2}\,S_0(2\pi\mu)
      \ear

It is worth to emphasize that an extreme example of the metric, for which the
relation $r(\rho)\ll L$ is satisfied, is that when $r(\rho)\equiv
r_0={\rm constant}$ and $f(\rho)\equiv f_0={\rm constant}$, i.e.,
        \beq\label{extreme_metric}
        ds^2= -f_0 dt^2+d\rho^2
            +r_0^2(d\theta^2+\sin^2\theta\, d\varphi^2).
        \eeq
To illustrate a case of the mertic with non-constant coefficients
obeying the condition $r(\rho)\ll L$ we may consider the following example:
$f(\rho)\equiv f_0={\rm constant}$ and
$r^2(\rho)=r_0^2[1+\alpha^2\ln(1+\rho^2/\rho_0^2)]$, where $\alpha$ and
$\rho_0$ are some parameters;
the relation $r(\rho)\ll L$ is satisfied provided $\alpha r_0/\rho_0\ll 1$.


Let, in addition to $r(\rho)\ll L$, the relation $r/\rc\gg 1$ be
satisfied. Stress that it assumes that $m\not=0$ and so $\mds=m$. In this
case $\mu\approx mr$ and we can simplify Eq. \Ref{phi2B} by using the
asymptotical form \Ref{appS} for $S_0(2\pi\mu)$ and taking into account that
$\ln(\mu^2/(\mds r)^2)=\ln(1+2\delta/(mr)^2)\approx
2\delta/(mr)^2-2\delta^2/(mr)^4$:
     \bear
     4\pi^2 m^2\phisq_\ren &=&\frac{1}{2 r^4}
      \left( { \xi^2 -\frac{\xi}{3}+\frac{1}{30} } \right).
     \ear
It is easily to see that the last expression is positive for all values
$\xi$.

Now consider the opposite relation $mr=r/\rc \ll 1$. Note that it
permits in particular the massless case $m=0$. In this case
$\mu\approx(2\delta)^{1/2}$ the expression \Ref{phi2B} for $\phisq$ can
approximately be rewritten as
      \bear
      4\pi^2\phisq_{\ren} =
      \frac{1}{2r^2}\left(\xi-\frac16\right)
      \ln\left(\frac{2\delta}{\mds^2 r^2}\right)
      -\frac{\delta}{2r^2}
      \left[1+4 S_0\left(2\pi\sqrt{2\delta}\right)\right].
      \ear
In particular, for conformal coupling $\xi=1/6$ ($\delta=1/24$) we obtain
$S_0(3^{-1/2}\pi) \approx -0.0596$ and
      \bear
      4\pi^2\phisq_{\ren}=-\frac{0.0159}{r^2}.
      \ear

Finally stress that the obtained expression \Ref{phi2B} for $\phisq$ is not
approximate but {\em exact} in the $R^2 \times S^2$ spacetime with the metric
\Ref{extreme_metric}.



\section*{Acknowledgment}

This work was supported by the Russian Foundation for Basic
Research grant No 99-02-17941.

\section*{Appendix: Asymptotics for $S_n(2\pi\mu)$ and $N_n^m(\mu)$}

\setcounter{equation}{0}
\renewcommand{\theequation}{A\arabic{equation}}
In this appendix we obtain asymptotics at large values of argument for the
functions $S_n(2\pi\mu)$ and $N_n^m(\mu)$. Consider the functions
$S_n(2\pi\mu)$ defined by Eqs. (\ref{appS}). Making the substitution
$y=2\pi\mu x$ in Eqs. (\ref{appS}) we find
        \beq\label{appS}
        \begin{array}{l}
        \dst
        S_0(2\pi\mu)=\lambda^2\int_0^\infty
        \frac{y\ln|1-\lambda^2 y^2|}
        {e^{y}+1}\,dy, \\[12pt]
        \dst
        S_n(2\pi\mu)=\int_0^\infty dy \ln|1-\lambda^2 y^2| \,
        \frac{d}{dy}
        \left(\frac1y\frac{d}{dy}\right)^{n-1}
        \frac{y^{2(n-1)}}{e^{y}+1},
        \end{array}
        \eeq
where $\lambda=(2\pi\mu)^{-1}$ and $\mu=(m^2r^2+2\delta)^{1/2}$. Note that
the integrands in Eqs. (\ref{appS}) contain exponential functions and are
exponentially decreasing at large values of $y$. Hence the main contribution
into the integrals is provided with values of integrands in the region $0 < y
< 1$. We are interesting in the case $\mu \gg 1$. In this case
$\lambda \ll 1$ and $\lambda y \ll 1$ if $0<y<1$. Now we may use the
asymptotical formula $\ln(1-\lambda^2 y^2)= -\lambda^2 y^2-\frac12 \lambda^4
y^4-\frac13 \lambda^6 y^6 +{\rm O}(\lambda^8 y^8)$. Substituting this into
Eqs. (\ref{appS}) and integrating gives
        \bear\label{asymS}
        &&S_0(2\pi\mu)=-\frac{7}{1920(mr)^4}+{\rm O}((mr)^{-6}), \qquad
        S_1(2\pi\mu)=\frac{1}{24(mr)^2}+{\rm O}((mr)^{-4}), \quad
        \nonumber \\
        &&S_2(2\pi\mu)={\rm O}((mr)^{-4}), \qquad
        S_3(2\pi\mu)={\rm O}((mr)^{-6}).
        \ear
Consider now the functions $N_n^m(\mu)$:
        \bear
        N_n^m(\mu)=\sum_{l=0}^\infty
        \frac{(l+\frac12)^m}{[\mu^2+(l+\frac12)^2]^n},
        \ear
where $n=2,3,4,\dots$ and $m=1,3,\dots,2n-3$. It is easily to check that
they obey the following recurrent formula:
       \bear\label{rel1}
       N_n^m(\mu)=N_{n-1}^{m-2}(\mu)-\mu^2 N_n^{m-2}(\mu),
       \ear
and in addition,
       \bear\label{rel2}
       N_n^1(\mu)=\frac{(-1)^{n+1}}{2^n (n+1)!} \,
       \left(\frac1\mu\frac{d}{d\mu}\right)^n N_2^1(\mu).
       \ear
Using these formulas, one can express any function $N_n^m(\mu)$ via
$N_2^1(\mu)$. The function $N_2^1(\mu)$ can be represented as follows [see
Ref. \cite{PBM}, Eq. (5.1.26.23)]:
        \bear
        &&N_2^1(\mu)=-\frac{i}{4\mu}\left[\psi'\left(\frac12-i\mu\right)
        -\psi'\left(\frac12+i\mu\right)\right],
        \ear
where $\psi(z)=\Gamma'(z)/\Gamma(z)$ is the digamma function.
Using the asymptotical properties of the digamma function (see, for example,
Ref. \cite{Abram}) and taking into account Eqs. \Ref{rel1}, \Ref{rel2} we
can obtain the following asymptotic for large values of $\mu$:
     \bear \label{asymN}
     &&N^1_2(\mu)=\frac{1}{2(mr)^2}
      -\frac{\xi-1/6}{(mr)^4}+{\rm O}((mr)^{-6}), \quad
     N^1_3(\mu)=\frac{1}{4(mr)^4}
     -\frac{\xi-1/6}{(mr)^6}+{\rm O}((mr)^{-8}),
\nn  &&N^3_3(\mu)=\frac{1}{4(mr)^2}
      -\frac{\delta}{2(mr)^4}+{\rm O}((mr)^{-6}), \quad
    N^1_4(\mu)=\frac{1}{6(mr)^6}+{\rm O}((mr)^{-8}),
\nn &&N^3_4(\mu)=\frac{1}{12(mr)^4}+{\rm O}((mr)^{-6}), \quad
    N^5_4(\mu)=\frac{1}{6(mr)^2}+{\rm O}((mr)^{-4}),
\nn &&N^1_5(\mu)=\frac{1}{8(mr)^8}+{\rm O}((mr)^{-10}), \quad
    N^3_5(\mu)=\frac{1}{24(mr)^6}+{\rm O}((mr)^{-8}),
\nn &&N^5_5(\mu)=\frac{1}{24(mr)^4}+{\rm O}((mr^{-6}), \quad
    N^7_5(\mu)=\frac{1}{8(mr)^{2}}+{\rm O}((mr)^{-4}),
\nn &&N^1_6(\mu)=\frac{1}{10(mr)^{10}}+{\rm O}((mr)^{-12}), \quad
    N^3_6(\mu)=\frac{1}{40(mr)^{8}}+{\rm O}((mr)^{-10}),
\nn &&N^5_6(\mu)=\frac{1}{60(mr)^{6}}+{\rm O}((mr)^{-8}), \quad
    N^7_6(\mu)=\frac{1}{40(mr)^{4}}+{\rm O}((mr)^{-6}),
\nn &&N^9_6(\mu)=\frac{1}{10(mr)^{2}}+{\rm O}((mr)^{-4}).
    \ear

\newpage

\begin{figure}[tbp]
\centerline{\epsffile{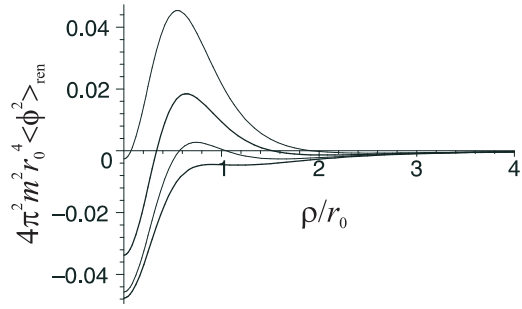}}
\caption{The curves in this figure display the the WKB approximation for
$\phisq$ \Ref{phi2A} calculated in the large $m$ limit in the Morris-Thorne
metric \Ref{metricMT}. From bottom to top the curves correspond to the values
$\xi=0,1/16,1/6,3/10$.}
\end{figure}

\begin{figure}[tbp]
\centerline{\epsffile{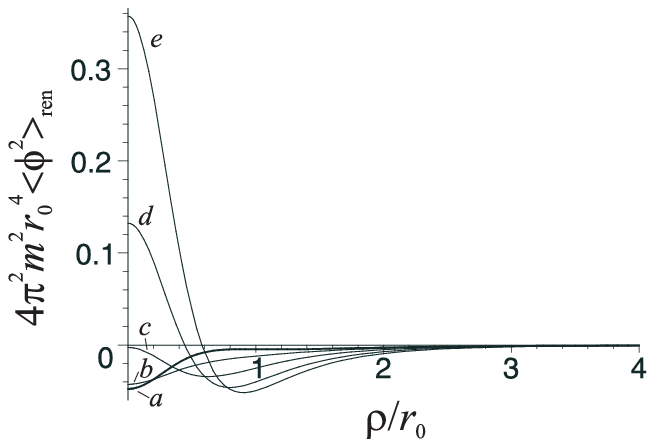}}
\caption{The curves in this figure display $\phisq$ calculated for negative
values of $\xi$. The labels $a,b,c,d,e$ mark curves which correspond to the
values $\xi=0,-0.1,-0.3,-0.6,-0.9$.}
\end{figure}


\begin{thebibliography}{99}
\bibitem{def}
Of course, this `definition' is not rigorous. For more details, see Ref.
\cite{VisserBook}. More recently Hochberg and Visser \cite{HV} gave a
precise, local definition of a (traversable) wormhole throat and
performed a detailed, pure local analysis of wormhole geometry. Also a fresh
discussion devoted to a notion of a wormhole can be found in Ref.
\cite{Krasnikov}
\bibitem{activ}
The reader can find some references on wormhole physics in the excelent
book by Visser \cite{VisserBook}
\end{thebibliography}
\end{document}